\documentclass[nofootinbib,amsfonts,prd,aps,12pt]{revtex4}

\usepackage[T1]{fontenc}

\usepackage{graphicx}
\usepackage{amsmath,amssymb}
\usepackage{hyperref}
\usepackage{graphicx}
\usepackage{subfigure}

\begin{document}

\title{Schr\"odinger-like quantum dynamics in loop quantized black holes} 
\author{Rodolfo Gambini}
\affiliation{Instituto de F\'{i}sica, Facultad de Ciencias, Igu\'a 4225, esq.\ Mataojo, Montevideo, Uruguay}
\author{Javier Olmedo}
\affiliation{Department of Physics and Astronomy, Louisiana State University, Baton Rouge, LA 70803-4001}
\author{Jorge Pullin}
\address{Department of Physics and Astronomy, Louisiana State University, Baton Rouge, LA 70803-4001}

\begin{abstract}
We show, following a previous quantization of a vacuum spherically symmetric spacetime carried out in Ref. \cite{gop}, that this setting admits a Schr\"odinger-like picture. More precisely, the technique adopted there for the definition of parametrized Dirac observables (that codify local information of the quantum theory) can be extended in order to accommodate different pictures. In this new picture, the quantum states are parametrized in terms of suitable gauge parameters and the observables constructed out of the kinematical ones on this space of parametrized states.
\end{abstract}

\maketitle

\section{Introduction}

Loop quantum gravity is a promising approach for the quantization of general relativity~\cite{lqg}. It emerged as a nonpertubative, background independent, canonical quantization of Einstien's theory. This quantization program is mathematically consistent, but its physical predictions (as well as its semiclassical limit) have not been fully understood, yet. Nevertheless, its quantization techniques have been successfully applied to several minisuperspace models and midisuperspace scenarios \cite{lqc}. The common shared feature is the existence of a semiclassical limit reproducing general relativity as well as the singularity is replaced by a region with large curvature \cite{aps,effec-dyn,effec-dyn2,gp-lett,gop,shell}. 

One of the outstanding situations that is receiving important attention nowadays concerns spherically symmetric gravity. In vacuum, Einstein's gravity describes in a very good approximation the fundamental physics of astronomical black holes. As these theoretical models were studied in depth, several fundamental questions arose, like what is the true physics behind the classical singularity or the nature of very interesting phenomena like Hawking radiation, black hole evaporation and eventually the information loss paradox. Within loop quantum gravity, it has been possible to answer some of these questions. For instance, the resolution of the singularity and the replacement by a region of high curvature has been observed by many authors \cite{bojo,ash-bojo,bs,bv,cgp,cgp2,bv2,chiou0,chiou,gp-lett,djp,cs}, as well as several aspects of Hawking radiation have been studied incorporating quantum gravity corrections \cite{gp-semi}. 

Recently, it was possible to find a full quantization of these vacuum settings in Ref. \cite{gp-lett}. However, the strategy followed there assumed periodicity for one of the variables of the model (a minor restriction that allowed to explicitly integrate the model). It also adopted a specific form for the quantum scalar constraint that is not obviously applicable in presence of matter. In a subsequent publication \cite{gop} the quantization was extended in order to deal with the previous questions. However, several aspects remain not fully understood. Concretely, the solutions to the scalar constraint were obtained by means of group averaging techniques \cite{raq}. It requires selfadjointness of the constraint operator. Assuming that the constraint fulfills that condition, one can simplify calculations by diagonalizing some geometrical operators. However, a detailed study of the spectrum requires numerical tools that are currently under development. Eventually, one should study the dynamics of semiclassical states in order to establish the semiclassical regime of the theory. 

Among the different strategies to achieve this goal, we will present in this manuscript a particularly attractive one. It is in parallel with the description commonly adopted in loop quantum cosmology. There, the solutions to the Hamiltonian constraint are computed (explicitly or numerically) after a suitable choice of an internal physical clock. Then, the solutions are equipped with Hilbert space structure and the observables defined as evolving constants of the motion by means of suitable kinematical operators. Here, we will provide a consistent description, assuming that the solutions to the constraint are known (they are currently under study but beyond the scope of this manuscript). We will also adopt here the group averaging techniques. The quantum scalar constraint is a symmetric operator on the kinematicl Hilbert space and we will assume that it is selfadjoint. We then parametrize the physical states (instead of the observables like in Refs. \cite{gp-lett,gop}) and define the basic physical operators of the model by means of the kinematical ones projected on the physical Hilbert space. This picture is particularly well adapted in a situation where the solutions to the constraint can be computed as well as a physical inner product is available. Finally, we comment on the relation between this picture (that we call Schr\"odinger-like picture) and the one in terms of parametrized observables (or Heisenberg-like picture) of Refs. \cite{gp-lett,gop}.

This manuscript is organized as follows. In Sec. \ref{sec:class-syst} we provide the classical setting that will be studied together with the Abelianization procedure for the Hamiltonian constraint. Sec. \ref{sec:kinemat} incorporates the kinematical representation adopted in this paper for the quantization of the model. We represent the scalar constraint in Sec. \ref{sec:scalar-const} and we study its solutions in Sec. \ref{sec:solut}. Finally, in Sec. \ref{sec:physical} we provide the physical picture of the model within the Schr\"odinger-like representation. Sec. \ref{sec:concl} is devoted to the conclusions.

\section{Classical system}\label{sec:class-syst}

\subsection{Kinematics}

Spherically symmetric gravity has been studied in Ashtekar-Barbero variables in Refs. \cite{bojo,bs} (see Refs. \cite{bencorteit,bengtsson,bojo2} for a general discussion). In summary, after the reduction, it is common to carry out a series of canonical transformations such that one is able to separate those fields that commute with the Gauss constraint (gauge invariant fields) from those which are pure gauge degrees of freedom. The Gauss constraint is well understood at the quantum level, and it is not relevant for all practical purposes in this manuscript. So, we will fix it at the classical level. 

The gauge invariant variables span the phase space, such that
\begin{align}\nonumber\label{eq:poiss}
&\{K_x(x),E^x(\tilde x)\}=G\delta(x-\tilde x),\\
&\{{K}_\varphi(x),E^\varphi(\tilde x)\}=G\delta(x-\tilde x),
\end{align} 
where $G$ is the Newton constant. The geometrical role of $E^x$ and $E^\varphi$ can be read from the spatial metric 
\begin{equation}
ds^2=\frac{(E^\varphi)^2}{|E^x|}dx^2+|E^x| (d\theta^2+\sin^2\theta d\phi^2),
\end{equation}
while $K_x$ and $K_\varphi$ represent the components of the extrinsic curvature in triadic form. It is worth commenting that for the component in the radial direction we have introduced a factor two for convenience following the notation of Refs. \cite{gop,gp-lett} (in other works like Refs. \cite{bs,cgp,chiou} this factor two is not considered).

The dynamics of the system is ruled by the Hamiltonian 
\begin{equation}\label{eq:total-ham}
H_T=\int dx (NH+N_rH_r),
\end{equation}
that is a linear combination of
the diffeomorphism and scalar constraints
\begin{subequations}
	\begin{align}
	& H_r:=G^{-1}[E^\varphi K_\varphi'-(E^x)' K_x]\,,\label{eq:difeo}\\ \nonumber
	&H :=G^{-1}\left\{\frac{\left[(E^x)'\right]^2}{8\sqrt{E^x}E^\varphi}
	-\frac{E^\varphi}{2\sqrt{E^x}} - 2 K_\varphi \sqrt{E^x} K_x  
	-\frac{E^\varphi K_\varphi^2}{2 \sqrt{E^x}}\right.\\
	&\left.-\frac{\sqrt{E^x}(E^x)' (E^\varphi)'}{2 (E^\varphi)^2} +
	\frac{\sqrt{E^x} (E^x)''}{2 E^\varphi}\right\}\,,\label{eq:scalar1}
	\end{align}
\end{subequations}
respectively. One can easily check that the constraint algebra is
\begin{subequations}
	\begin{align}
	&\{H_r(N_r),H_r(\tilde N_r)\}=H_r(N_r\tilde N_r'-N_r'\tilde N_r),\\
	&\{H(N),H_r(N_r)\}=H(N_r N'),\\
	&\{H(N),H(\tilde N)\}=H_r\left(\frac{E^x}{(E^\varphi)^2}\left[N\tilde N'-N' \tilde N\right]\right).
	\end{align}
\end{subequations}
We observe that it is equipped with structure functions (like in the general theory), with the
ensuing difficulties for achieving a consistent quantization \cite{haji-kuch}.

\subsection{Weak Dirac observables: black hole mass}

It is worth commenting that the classical theory possesses two weak Dirac observables (only one of them is linearly independent on-shell). They have a Poisson bracket with the Hamiltonian that vanishes on-shell. Let us define the phase space functions of weight density zero
\begin{align}
&{\cal M}(x):=\frac{2E^x\sqrt{E^x} K_x K_\varphi}{GE^\varphi}+\frac{\sqrt{E^x}E^x(E^x)' (E^\varphi)'}{2G (E^\varphi)^3} -
\frac{\sqrt{E^x} E^x(E^x)''}{2 G(E^\varphi)^2},\\
&\tilde {\cal M}(x):=-\frac{1}{2G}\sqrt{E^x}(1+K_\varphi^2) +\frac{\sqrt{E^x}[(E^x)']^2}{8G(E^\varphi)^{2}}.
\end{align}
After some calculations, it is possible to see that, on-shell, one would have $\dot{{\cal M}}(x)= 0$. The calculations are rather lengthly but not complicated, so we will not show them here. We conclude that ${\cal M}(x)$ is a weak Dirac observable of the model. This observable is just the ADM mass ${{\cal M}}(x)= M$ on-shell. 

Regarding $\tilde {\cal M}(x)$, one can easily see that $\tilde {\cal M}(x)={\cal M}(x)-E^x H/E^\varphi$ is in fact a linear combination of the weak observable ${\cal M}(x)$ and the scalar constraint $H$. It is not difficult to convince oneself that ${\dot{\tilde{\cal M}}}(x)= 0$ on the constraint surface. So $\tilde {\cal M}(x)$ is another weak observable. But let us notice that $\tilde {\cal M}(x)={\cal M}(x)$ on-shell, so there is only one physical degree of freedom: the mass of the spacetime.

\subsection{Abelianization}

To proceed with the quantization by means of a Dirac quantization approach, we still have to overcome an important obstacle: the constraint algebra possesses structure functions. This fact does not impede the obtaining of a consistent quantum theory. But one has to work with a non-selfadjoint set of constraints \cite{komar1,komar2,haji-kuch,brahma} in the most favorable case (to the knowledge of the authors). If one adopts a quantization following the ideas of loop quantum gravity, the situation becomes more interesting since one has to deal with additional anomalies with respect to a standard representation (see for instance Refs. \cite{brahma,bojo3,bb,bbr}).

One strategy that has been recently adopted for the quantization of spherically symmetric vacuum spacetimes \cite{gp-lett,gop} as well as in polarized vacuum Gowdy models with local rotational symmetry \cite{gowdy-LRS} consists in redefining the scalar constraint in a suitable way such that the new Hamiltonian constraint has an Abelian algebra with itself. We will adopt this strategy in this manuscript.

Let us begin with a brief description of the procedure. It is well known from totally constrained theories that, given a first class system (like the one under study), it is possible to find a new set of constraints such that the new ones are written as a linear combination of the former and the Poisson brackets among them vanish \cite{hennteit}. However, this Abelianization is not commonly employed since the new set of constraints usually have a complicated functional form of the phase space functions with respect to the original set of constraints. Furthermore, as it was discussed in Ref. \cite{hennteit}, the Abelianization holds only on a region of phase space that, in general, does not cover the whole constraint surface.

For our purposes, we will adopt a redefinition of only the scalar constraint, leaving the diffeomorphism one unaltered. The reason is two fold. On the one hand, the Poisson algebra of the Hamiltonian constraint with itself shows structure functions. They potentially difficult the adoption of a Dirac quantization approach. It seems advantageous then to adopt the corresponding Abelianization assuming that the physical picture it provides incorporates the main quantum geometry corrections. The diffeomorphism constraint, on the other hand, will remain unaltered since it generates the classical finite diffeomorphisms that are implemented by loop quantum gravity in a robust way (by averaging the graphs with respect to such set of diffeomorphims). 

More concretely, let us consider the following linear combination of constraints 
\begin{equation}\label{eq:hnew}
H_{\rm new} :=\frac{\left(E^x\right)'}{E^\varphi}H-2\frac{\sqrt{E^x}}{E^\varphi}K_\varphi H_r=\frac{1}{G}\left[ \sqrt{E^x}\left(1-\frac{[(E^x)']^2 }{4 (E^\varphi)^2}+K_\varphi^2\right)\right]',
\end{equation}
that defines the new scalar constraint $H_{\rm new}$. At the level of the action, this redefinition amounts to absorb part of the constraints in a new lapse and shift functions given by the relation
\begin{equation}
N^{\rm new}_r:= N_r -2 N\frac{K_\varphi\sqrt{E^x}}{\left(E^x\right)'},\quad N_{\rm new} := N \frac{E^\varphi}{\left(E^x\right)'}.
\end{equation}
We see what can go wrong with this redefinition: those situations where $\left(E^x\right)'=0$. It is typical to adopt a gauge fixing condition of the form of $\left(E^x\right)'=0$ for studying the interior of the black hole (Kantowski-Sachs spacetime) ---see Ref. \cite{ash-bojo,cgp2}---. In this case, we think that either one should convince oneself that $\left(E^x\right)'=0$ is a bad gauge fixing condition for the new set of constraints, and then consider a different coordinate system that allows one to penetrate into the interior of the black hole, or one should work with the original set of classical constraints for this particular case (the reduced quantum theory can be completed since the structure functions vanish \cite{cgp2}). It is worth commenting that in the quantum theory factor ordering ambiguities can be still used in order to face this aspect. For instance, it is possible to represent the operator $\widehat{[\left(E^x\right)']}^{-\epsilon}$ with any $\epsilon>0$ in such a way that when $\left(\hat E^x\right)'=0$ on a given state, the operator $\widehat{[\left(E^x\right)']}^{-\epsilon}$ is still finite and well defined (see Ref. \cite{inv-thiem}). Of course, it would probably involve the emergence of important quantum effects at regimes where they would be unexpected. 

Eventually, it is more natural to work with an integrated version of $H_{\rm new}$. However, at the level of the action, this requires integrating by parts the new Hamiltonian constraint times the new lapse. In order to carry out this integration consistently, it is necessary to consider suitable boundary conditions. They have been extendedly studied (see for instance Ref. \cite{kuchar}).  
Taking into account the new lapse $\tilde N :=- N_{\rm new}'\sqrt{E^x}(E^\varphi)^{-1}$, we get the scalar constraint
 \begin{equation}
\tilde H(\tilde N) :=\frac{1}{G}\int dx \tilde N
 E^\varphi\bigg[ K_\varphi^2-\frac{[(E^x)']^2}{4
 	(E^\varphi)^2}+\left(1-\frac{2 G M}{\sqrt{E^x}}\right)\bigg],
 \end{equation}
where $M$ is the mass of the black hole that must be considered as an additional degree of freedom in order to have a well-posed variational problem.
 
The final action would be
\begin{equation}
S=\int dt \left[M\dot \tau+\int dx G^{-1}\left(E^\varphi\dot K_\varphi+E^x\dot K_x\right)-\tilde H(\tilde N)-H_r(N_r)\right],
\end{equation}
such that $\tau$ is the canonically conjugated variable of $M$.

\section{Quantization: kinematical structure}\label{sec:kinemat}

Let us continue with the quantization of the classical theory. On the one hand, for the mass of the black hole one can adopt a standard
description $L^2(\mathbb{R},dM)$. It is worth commenting that one can select an alternative representation for this degree of freedom (see for instance Ref. \cite{thiem-kast}). In this case, the measure $dM$ would be compatible with dilations of the mass instead of translations. A new observable $\epsilon=\pm1$ emerges corresponding to the sign of the classical mass. The restriction to the sector $\epsilon=1$ corresponds to $M>0$ and guarantees the positivity condition of $\hat M$. We will adhere to this representation, i.e. ${\cal H}_{\rm kin}^m=L^2(\mathbb{R}^+,dM)$. On the other hand, for the geometrical degrees of freedom we will consider the kinematical Hilbert space whose structure is inherited from  loop quantum gravity \cite{bs}. Here, the basic bricks are one-dimensional graphs formed by combinations of  holonomies of $su(2)$-connections along non-overlapping edges $e_j$ connecting vertices $v_j$. It is natural to associate the variable $K_x$ with edges
along the radial direction in the graph and the scalar $K_\varphi$ with vertices on it (point holonomies). Each edge will be connected with
another one by means of the corresponding vertex. A given graph can be written as 
\begin{align}
&\Psi_{g,\vec{k},\vec{\mu}}(K_x,K_\varphi) =\prod_{e_j\in g}
\exp\left(i \gamma k_{j} \int_{e_j} dx\,K_x(x)\right)
\prod_{v_j\in g}
\exp\left(i  \gamma \mu_{j} K_\varphi(v_j) \right),
\end{align}
where the label  $k_j\in\mathbb{Z}$ and $\mu_j\in\mathbb{R}$ are the valences associated with the edge $e_j$ and the vertex $v_j$, respectively. Usually one refers to them as the ``coloring'' of the edges and vertices. In the expression above, $\gamma$ is the Immirzi parameter \cite{lqg}.

Let us recall that the point holonomies are associated to variables that are not true connections. In this case, the closer kinematical structure to the full theory would be a representation in terms of quasiperiodic functions of these connection-like variables, as is usually done in loop quantum cosmology, whose kinematical Hilbert space is
$L^2(\mathbb{R}_{\rm Bohr},d\mu_{\rm Bohr})$, with
$\mathbb{R}_{\rm Bohr}$ the Bohr compactification of the real line and $d\mu_{\rm
	Bohr}$ the natural invariant measure under translations on that set. 

The kinematical Hilbert space ${\cal H}_{\rm kin}$, for a given graph $g$, turns out to be the tensor product
\begin{equation}
{\cal H}_{\rm kin}={\cal H}_{\rm kin}^M\otimes\left[\bigotimes_{j=1}^n\ell^2_j\otimes L_j^2(\mathbb{R}_{\rm Bohr},d\mu_{\rm Bohr})\right],
\end{equation}
with $n$ the number of vertices of the graph. ${\cal H}_{\rm kin}$ is endowed with the inner product
\begin{align}\label{eq:kin-inner-prod-qp}
\langle g,\vec{k},\vec{\mu},M
|g',\vec{k}',\vec{\mu}',M'
\rangle=\delta(M-M')\delta_{\vec{k},\vec{k}'}\delta_{\vec{\mu},\vec{\mu}'}\delta_{g,g'}\;.
\end{align}
Here $\ell^2_j$ means the Hilbert space of square summable functions corresponding to the holonomies of the connection $K_x$ in the radial direction.

The basic operators are represented as 
\begin{subequations}
\begin{align}
&{\hat{M} } |g,\vec{k},\vec{\mu},M\rangle
= M |g,\vec{k},\vec{\mu},M\rangle,\\
&{\hat{E}^x(x) } |g,\vec{k},\vec{\mu},M\rangle
= \gamma \ell_{\rm Pl}^2 k_j |g,\vec{k},\vec{\mu},M\rangle,
\\
&\hat{E}^\varphi(x) |g,\vec{k},\vec{\mu},M\rangle
= \gamma \ell_{\rm Pl}^2 \sum_{v_j} \delta\big(x-x_j\big)\mu_j 
|g,\vec{k},\vec{\mu},M\rangle,
\end{align}
\end{subequations}
where $k_j$ is either the valence of the edge on the point $x\in e_j$ or, if $x$ corresponds to a vertex $v_j$, the valence of the edge $e_j$. Besides,
$x_j$ is the position of the vertex $v_j$, with $j=1,2,\ldots$

It is worth commenting that the only connection component that is present in the scalar constraint, $K_\varphi(x)$, will be represented in terms of point holonomies of length $\rho(x)$. In order to visualize explicitly the action of these operators on kinematical states, let us notice first that
\begin{equation}
\hat N^\varphi_{\rho}(x) = \widehat{e^{i  \gamma \rho(x) K_\varphi(x)}}.
\end{equation}
Let us now write the kinematical states as
\begin{equation}\label{eq:ket-dec}
|g,\vec{k},\vec{\mu},M\rangle=\left(\bigotimes_{v_j\in g}|k_j\rangle\otimes|\mu_j\rangle\right)\otimes|M\rangle.
\end{equation}

If the operator $\hat N^\varphi_{\rho}(x)$ is evaluated at $x=x_j$, i.e., on a given vertex, its action on a ket $|\mu_j\rangle$ will be
\begin{equation}
\hat N^\varphi_{\rho_j}(x_j) |\mu_j\rangle
= |\mu_j+\rho_j\rangle,
\end{equation}
where $\rho_j=\rho(x_j)$. On the other hand, if $x$ is not on a vertex, for instance $x_{j}<x<x_{j+1}$, the action on $|g,\vec{k},\vec{\mu},M\rangle$ will be
\begin{equation}\label{eq:ket-dec2}
\hat N^\varphi_{\rho}(x)|g,\vec{k},\vec{\mu},M\rangle=\left(\bigotimes_{i=1}^j|k_i\rangle\otimes|\mu_i\rangle\right)\otimes\left(|k_j\rangle\otimes|\rho(x)\rangle\right)\otimes\left(\bigotimes_{i={j+1}}^n|k_i\rangle\otimes|\mu_i\rangle\right)\otimes|M\rangle.
\end{equation}
In this case, as we explicitly showed, the operator $N^\varphi_{\rho}(x)$ creates a new vertex at $x$ between $x_{j}$ and $x_{j+1}$.

\section{Representation of the Hamiltonian Constraint}\label{sec:scalar-const}

The quantum Hamiltonian constraint is similar to the one adopted in Ref. \cite{gop}. Concretely, it is given by
\begin{align}\label{eq:quant-scalar-constr}
&  \hat{H}(N)=\frac{1}{G}\int dx N(x)\hat P\left\{  
\hat\Theta
-\frac{1}{4}\widehat{\left[\frac{\left[({E}^x)'\right]^2}{{E}^\varphi}\right]} +\hat{E}^\varphi\left(1
-\frac{2 G \hat M}{\sqrt{\hat{E}^x}}\right) \right\}\hat P.
\end{align}
The operator $\hat P$ is defined as follows
\begin{equation}
\hat P|g,\vec{k},\vec{\mu},M\rangle=\prod_{v_j}{\rm sgn}(k_j){\rm sgn}(\mu_j)|g,\vec{k},\vec{\mu},M\rangle,
\end{equation}
such that ${\rm sgn}(x)$ is the standard sign function with ${\rm sgn}(0)=0$. Therefore, states with any $k_j=0$ and/or $\mu_j=0$ are annihilated by the constraint. We will therefore restrict the study to the orthogonal complement of these states, i.e., those states with nonvanishing $k_j$ and $\mu_j$. Then, the triads cannot vanish in this subspace. The consequence is that the classical singularity, as well as some of the coordinate singularities, will be cured.

The operator $\hat\Theta(x)$ acting on the kinematical states 
\begin{align}
&\hat\Theta(x)|g,\vec{k},\vec{\mu},M\rangle
= \sum_{v_j\in g} \delta(x-x_j) \hat\Omega_\varphi^2 (x_j)
|g,\vec{k},\vec{\mu},M\rangle,
\end{align}
is defined by means of the non-diagonal operator
\begin{align}
&\hat{\Omega}_\varphi (x_j)= \frac{1}{{8i\rho\gamma}}|\hat{E}^\varphi|^{1/4}\big[\widehat{{\rm sgn}(E^\varphi)}\big(\hat N^\varphi_{2\rho}-\hat N^\varphi_{-2\rho}\big)+\big(\hat N^\varphi_{2\rho}-\hat N^\varphi_{-2\rho}\big)\widehat{{\rm sgn}(E^\varphi)}\big]|\hat{E}^\varphi|^{1/4}\Big|_{x_j},
\end{align}
where
\begin{equation}
|\hat{E}^\varphi|^{1/4}(x_j)  |g,\vec{k},\vec{\mu},M\rangle
= (\gamma\ell_{\rm Pl}^{2}  |\mu_j|)^{1/4}
|g,\vec{k},\vec{\mu},M\rangle,
\end{equation}
\begin{equation}
\widehat{{\rm sgn}\big(E^\varphi(x_j)\big)}  |g,\vec{k},\vec{\mu},M\rangle
= {\rm sgn}(\mu_j)
|g,\vec{k},\vec{\mu},M\rangle,
\end{equation}
have been constructed by means of the spectral decomposition of $\hat{E}^\varphi$ on
${\cal H}_{\rm kin}$. Finally, we will consider 
\begin{align}
&\widehat{\left[\frac{\left[({E}^x)'\right]^2}{{E}^\varphi}\right]} |g,\vec{k},\vec{\mu},M\rangle
=\sum_{v_j\in g} \delta(x-x_j)\gamma\ell_{\rm Pl}^{2}(\Delta k_j)^2b^2_{\rho_j}(\mu_j)
|g,\vec{k},\vec{\mu},M\rangle.
\end{align}
where 
\begin{equation}
\Delta k_j=k_j-k_{j-1},
\end{equation}
and
\begin{equation}\label{eq:bmu}
b_{\rho_j}(\mu_j)=\frac{1}{\rho_j}(|\mu_j+\rho_j|^{1/2}-|\mu_j-\rho_j|^{1/2}),
\end{equation}
is defined by means of Thiemann's trick. Another useful choice of this function is to take the limit in which $\rho_j\to0$ that would be given by $b_0(\mu_j)=\mu_j^{-1/2}$ if $\mu_j\neq 0$, and $b_0(0)=0$.

The action of the constraint on the spin networks is
\begin{align}
&  \hat{H}(N) |g,\vec{k},\vec{\mu},M\rangle=\frac{1}{G}\sum_{v_j\in g} N(x_j)  \hat C_j|g,\vec{k},\vec{\mu},M\rangle.
\end{align}
If we recall the decomposition in Eq. \eqref{eq:ket-dec}, the action of the constraint $\hat C_j$ is given by
\begin{align}\nonumber 
& \hat C_j|\mu_j\rangle=f_0(\mu_j,k_j,k_{j-1},M) |\mu_j\rangle\\
&-f_+(\mu_j)|\mu_j+4\rho_j\rangle-f_-(\mu_j)|\mu_j-4\rho_j\rangle,
\end{align}
with
\begin{subequations}
\begin{align}\label{eq:fpm0}
&f_\pm(\mu_j)=\frac{\ell_{\rm Pl}^2}{64\gamma\rho_j^2}|\mu_j|^{1/4}|\mu_j\pm 2\rho_j|^{1/2}|\mu_j\pm 4\rho_j|^{1/4}s_{\pm}(\mu_j)s_{\pm}(\mu_j\pm 2\rho_j),\\\nonumber
&f_0(\mu_j,k_j,k_{j-1},M)=\frac{\ell_{\rm Pl}^2}{64\gamma\rho_j^2}\left[(|\mu_j||\mu_j+ 2\rho_j|)^{1/2}s_+(\mu_j)s_-(\mu_j+2\rho_j)
\right.\\\nonumber
&\left.+(|\mu_j||\mu_j- 2\rho_j|)^{1/2}s_-(\mu_j)s_+(\mu_j-2\rho_j)\right]+\gamma \ell_{\rm Pl}^2\mu_j\left(1-\frac{2GM}{\sqrt{\gamma\ell_{\rm Pl}^2|k_j|}}\right)\\\label{eq:f0}
&-\frac{\gamma\ell_{\rm Pl}^2}{4}(\Delta k_j)^2b^2_{\rho_j}(\mu_j),
\end{align}
\end{subequations}

This means that any solution to the constraint preserves the number of vertices or edges, has support on semilattices of the form $\mathcal{L}_{\varepsilon_{j}}=\{\mu_{j} | \mu_{j}=\varepsilon_{j}+4\rho_jm,\ m\in \mathbb{N},\ \varepsilon_{j}\in(0,4\rho_j]\}$, and preserves the sequences $\{k_j\}$.

We note that the constraints $\hat C_j$ are symmetric on ${\cal H}_{\rm kin}$ and unbounded. Therefore, a more detailed analysis is required in order to prove their selfadjointness, like a study of the spectrum of the set of constraints $\hat C_j$. In conclusion, we do not have a proof about their selfadjointness. Nevertheless, we will assume that this is the case in what follows.

\section{Solutions to the constraint}\label{sec:solut}

We can write the constraint equation in an explicit form. A general solution to the constraint $(\Psi_g|$ fulfills
\begin{equation}
\sum_{v_j\in g}(\Phi_g|N_j\hat{C_j}^\dagger=0,\quad\Leftrightarrow \quad (\Phi_g|\hat{C_j}^\dagger=0.
\end{equation}
Since they will be of the form 
\begin{equation}
(\Phi_g|=\int_{0}^\infty \!\!dM\sum_{\vec k}\sum_{\vec \mu}\langle g, \vec{k},\vec{\mu},M|\phi(\vec k,\vec{\mu},M),
\end{equation}
it is then natural to adopt the factorization 
\begin{equation}
\phi(\vec k,\vec{\mu},M)=\prod_{v_j}\phi_j(\mu_j), \quad \phi_j(\mu_j)=\phi_j(k_j,k_{j-1},\mu_j,M).
\end{equation}
One can easily see that the difference equation for each $\phi_j(\mu_j)$ is
\begin{align}\nonumber
&-f_+(\mu_j-4\rho_j)\phi_j(\mu_j-4\rho_j)-f_-(\mu_j+4\rho_j)\phi_j(\mu_j+4\rho_j)\\
&+f_0(k_j,k_{j-1},\mu_j,M) \phi_j(\mu_j)=0,
\end{align}
with the functions $f_\pm(\mu_j)$ and $f_0(k_j,k_{j-1},\mu_j,M)$ defined in Eqs. \eqref{eq:fpm0} and \eqref{eq:f0}. Let us recall, as it was first noticed in Ref. \cite{gop}, that the solutions to this difference equation have different asymptotic behaviors  depending on the sign of 
\begin{equation}\label{eq:F}
F_j=\left(1-\frac{2GM}{\sqrt{\gamma\ell_{\rm Pl}^2|k_j|}}\right).
\end{equation}
If it is positive, the quantum constraints $\hat C_j$ can be written in a separable form where one of the geometrical operators has normalizable eigenfunctions (discrete spectrum). Similarly, if $F_j<0$, then $\hat C_j$, which can again be written in a separable form, is determined by a difference operator with normalizable eigenfunctions in the generalized sense (the spectrum is continuous). Let us see this in more detail.

\subsection{Continuous spectrum for $F_j<0$}

Let us consider those vertices where $k_j$ and $M$ are such that $F_j<0$. In this case, we can write the difference equation on a given vertex in a suitable separable form if we solve for the solutions 
\begin{equation}
	\phi^{\rm cnt}_j(\mu_j)=\hat \mu_j^{1/2}\phi_j(\mu_j),
\end{equation}
where we must recall that $\phi_j(\mu_j)$ is the solution of the original constraint in the vertex $v_j$ and that the states with $\mu_j=0$ have been decoupled (this transformation is invertible). In this case, the constraint at a given vertex corresponding to $\phi^{\rm cnt}_j(\mu_j)$ is given by 
\begin{equation}\label{eq:in-eigen-eq}
\hat \mu_j^{-1/2}\hat C_j\hat \mu_j^{-1/2}=\hat{\cal C}_j^{\rm cnt}+\left(1-\frac{2GM}{\sqrt{\gamma\ell_{\rm Pl}^2|k_j|}}\right),
\end{equation}
where $\hat{\cal C}_j^{\rm cnt}$ is a difference operator for each $j$. We can determine its eigenvalues and the corresponding eigenfunctions by means of the difference equation
\begin{equation}
\hat{\cal C}_j^{\rm cnt}|\phi^{\rm cnt}_{\omega_j}\rangle=\ell_{\rm Pl}^2\omega_j|\phi^{\rm cnt}_{\omega_j}\rangle,
\end{equation}
where we have introduced the factor $\ell_{\rm Pl}^2$ for convenience (so that $\omega_j$ is dimensionless). The eigenfunctions fulfill the normalization condition 
\begin{equation}
\langle \phi^{\rm in}_{\omega_j}|\phi^{\rm in}_{\omega'_j}\rangle=\delta\left(\sqrt{\omega_j}-\sqrt{\omega_j'}\right),
\end{equation}
for $\omega_j>0$ \cite{gop}. As we mentioned, we have not analyzed with sufficient detail the eigenfunctions of the difference equation for negative eigenvalues as well as complex ones. Since we expect that $\hat{\cal C}_j^{\rm cnt}$ will be selfadjoint, no complex eigenvalues will be found. Besides, since the physically relevant eigenvalues correspond to the positive ones, we will restrict the study to them  in this manuscript, neglecting so the eigenstates corresponding to negative eigenvalues.

The constraint equation in this basis, for each vertex, takes the simple form
\begin{equation}\label{eq:const-eq}
\omega_j+\left(1-\frac{2GM}{\sqrt{\gamma\ell_{\rm Pl}^2|k_j|}}\right)=0.
\end{equation}

It is worth commenting that the eigenvalues in the continuum part of the spectrum cannot be arbitrarily large since the eigenfunctions oscillate with a maximum frequency dictated by the step of the lattice in $\mu_j$. We can estimate this bound by recalling that, in a lattice of step $\Delta$, the maximum frequency is given by $\omega_{\rm max}^2=\frac{4}{\Delta^2}$. In our case, since $\Delta=4\rho_j$, we conclude that only $\omega_j\lesssim \frac{1}{4\gamma \rho_j^{2}}$ will propagate.\footnote{It is worth commenting that there could exist normalizable solutions for eigenvalues larger than $\omega_{\rm max}$, but we have not yet studied this possibility in detail.} It is not completely clear to us what the physical consequences are of this cutoff, but it seems that it is the analog of the quantum bounce in Kantowski-Sachs models \cite{cgp2,ash-bojo,bv,bv2,chiou0,djp}.

\subsection{Discrete spectrum for $F_j>0$}

In this section we will solve the constraint operator for those vertices where $F_j>0$. In this case, the difference equation for the solutions can be written again in a separable form if we instead study the difference equation of $\phi^{\rm dcr}_j(\mu_j)$ with
\begin{equation}\label{eq:scaling-out}
\phi^{\rm dcr}_j(\mu_j)=\hat b_{\rho_j}(\mu_j)\phi_j(\mu_j),
\end{equation}
where  $b_{\rho_j}(\mu_j)$ is given in \eqref{eq:bmu}. In this situation,
the constraint equation on a vertex $v_j$ and for the solution $\phi^{\rm dcr}_j(\mu_j)$ admits a separation of the form
\begin{equation}\label{eq:const-ext}
\hat b_{\rho_j}(\mu_j)^{-1}\hat C_j\hat b_{\rho_j}(\mu_j)^{-1}=\hat{\cal C}_j^{\rm dcr}-\frac{\gamma \ell_{\rm Pl}^2}{4}(\Delta k_j)^2,
\end{equation}
Here $\hat{\cal C}_j^{\rm dcr}$ is a difference operator, for each $j$, whose spectral decomposition can be carried out by solving the eigenvalue problem
\begin{equation}\label{eq:dcr-eigsys}
\hat{\cal C}_j^{\rm dcr}|\phi^{\rm dcr}_{\lambda_j}\rangle=\ell_{\rm Pl}^2\lambda_j|\phi^{\rm dcr}_{\lambda_j}\rangle.
\end{equation}
Again, we have introduced a factor $\ell_{\rm Pl}^2$ in order to make $\lambda_j$ dimensionless. The corresponding eigenfunctions are now normalized to \cite{gop}
\begin{equation}
\langle \phi^{\rm out}_{\lambda_n(\varepsilon_j)}|\phi^{\rm out}_{\lambda_{n'}(\varepsilon_j')}\rangle=\delta_{nn'}\delta_{jj'}.
\end{equation}

We obtain the eigenfunctions following the ideas of Ref.~\cite{cfrw} (where the homogeneous
constraint equation is analogous to ours at each vertex $v_j$).  
The spectrum of the corresponding difference operator turns out to be discrete owing to the behavior of its eigenfunctions at $\mu_j\sim \varepsilon_j$ and $\mu_j\to \infty$. At $\mu_j\sim \varepsilon_j$ the eigenfunctions cannot oscillate infinitely before reaching $\varepsilon_j$. Therefore, we expect that $\lambda_n$ will belong to a countable set (for each vertex $v_j$), which can be determined numerically. Our numerical investigations show that $\lambda_{n}$ depends on $\varepsilon_j\in(0,4\rho_j]$, as well as on $F_j$, i.e., on $k_j$ and $M$ by means of Eq.~\eqref{eq:F}.
 
Therefore, the constraint equation (up to a factor $ \ell_{\rm Pl}^2$) in this basis takes the algebraic form
\begin{equation}\label{eq:algeb-const}
\lambda_n(\varepsilon_j)-\frac{\gamma}{4}(\Delta k_j)^2=0.
\end{equation}
At this point, it is interesting to ask which are the physical consequences if Eq. \eqref{eq:algeb-const} must be fulfilled. Let us notice that there seems to be some tension since the second term will be the square of an integer (times $\gamma$ over four), while $\lambda_n(\varepsilon_j)$ is discrete once $F_j$ and $\varepsilon_j$ are fixed. We can consider two strategies. The first one is to choose a value for $F_j$ and then check the dependence of the eigenvalues with respect to the parameter $\varepsilon_j$, that can take any real value in $(0,4\rho_j]$, and see whether it is possible to find a suitable choice of $n$ and $\varepsilon_j$  such that Eq. \eqref{eq:algeb-const} is fulfilled. This is feasible as we have seen from our numerical studies. The physical states in general are superpositions of different $\vec{k}$ and $M$. That implies that we will be superposing states on different semilattices. Therefore, the discretization of the triad $\hat E^\varphi$ will be softened. 

On the other hand, the second possibility is to restrict the numerical study to one semilattice given by $\varepsilon_j$ and check which values of $F_j$ are compatible with Eq. \eqref{eq:algeb-const}. This possibility seems to be too restrictive. In the asymptotic region $k_j\to\infty$, where $F_j\simeq 1$, the possible values of $\Delta k_j$ seem to be strongly restricted. Therefore, the most plausible situation for solving the mentioned tension is to allow for superpositions of different $\varepsilon_j$.

Let us comment that, as in the previous subsection, we have not studied the eigenfunctions of $\hat{\cal C}_j^{\rm dcr}$ corresponding to negative or even  complex eigenvalues. Due to the form of Eq. \eqref{eq:algeb-const} the negative ones are not relevant, while we do not expect to find solutions for complex eigenvalues. Although a more careful study will be a matter of future research.

\section{Physical Hilbert space and observables: Schr\"odinger-like picture}\label{sec:physical}

We have been able to find a basis of states in the kinematical Hilbert space where the constraint equations become diagonal. The solutions to the scalar constraint can be constructed together with a suitable inner product by means of group averaging with the scalar constraint \cite{gop,raq}. We will present in this section a physical description that resembles the Schr\"odinger picture in quantum mechanics. It involves parametrized states instead of parametrized observables like in Refs. \cite{gp-lett,gop,gp-semi}.
The solutions to the scalar constraint take the form
\begin{equation}\label{eq:solution}
(\Psi^{C}_g|=\int_0^{\infty} dM\left(\bigotimes_j\left[\sum_{k_j}\psi(M,k_j)\langle\phi(k_j,M)|\otimes\langle k_j| \right]\right)\otimes \langle M|,
\end{equation}
where the bra $\langle \phi(\vec{k},M) |$ indicates that each of the eigenstates described in the previous section are such that the corresponding eigenvalues satisfy either Eq. \eqref{eq:const-eq} or Eq. \eqref{eq:algeb-const}. 

Finally, one can find the (spatially) diffeomorphism invariant states by group averaging the solutions to the Hamiltonian constraint with respect to the group of diffeomorphism \cite{raq-lqg}. The final physical states $\langle\Psi_{\rm Phys}|$ are superpositions of graphs such that the vertices are in all possible positions of the original one-dimensional manifold but preserving the order of the vertices and the orientation of the edges (see App. A of Ref. \cite{bbr} for additional details). 

This construction yields a well defined inner product 
\begin{equation}\label{eq:undep-inner}
\langle\Psi_{\rm Phys}|\Psi_{\rm Kin}\rangle=\int_0^{\infty} dM\sum_{\vec{k}}|\psi(M,\vec{k})|^2.
\end{equation}

Let us now construct a parametrization for the states with respect to suitable parameter functions. This is analogous to the parametrization carried out usually in quantum cosmology where one identifies a suitable phase space function, declares it as an internal time (gauge parameter), and parametrices the physical system in terms of this physical clock. We may notice first that the solutions of the constraints $\langle \Psi_{\rm Phys}|$ are superpositions of graphs with vertices in all possible positions within the equivalence class of graphs related by a spatial diffeomorphism. Then, we can parametrize these states as follows. Consider an arbitrary kinematical state $|\Psi\rangle_{\rm kin}$. They all are defined on a given one-dimensional manifold with the vertices located at concrete positions on this manifold. So that, in the inner product in Eq. \eqref{eq:undep-inner}, only the projection of $\langle \Psi_{\rm Phys}|$ on $|\Psi\rangle_{\rm kin}$ will contribute. However, let us remark that we can characterize all the kinematical states related to $|\Psi\rangle_{\rm kin}$ by means of a spatial diffeomorphism $x\to z(x)$ as $|\Psi\rangle_{\rm kin}\to|\Psi\left(z\right)\rangle_{\rm kin}$. In the same way, we can define analogously the family of all physical states projected on these kinematical states as $\langle\Psi_{\rm Phys}\left(z\right)|$. So, $\langle\Psi_{\rm Phys}|$ can be understood as the sum in $z$ of the states $\langle\Psi_{\rm Phys}\left(z\right)|$. This gives a suitable parametrization with respect to the choice of spatial coordinates.

Finally, let us parametrize the states for a particular choice of time function. For simplicity we will choose the connection $K_\varphi(x)$.\footnote{The connection has already been employed as an internal time in Loop Quantum Cosmology. See for instance Ref. \cite{bianchi}.} We must recall that only the exponentiation of this phase space function is well defined as an operator in the quantum theory. We then project the parametrized solutions to the constraint $\langle \Psi_{\rm Phys}\left(z\right)|$ on the ket $|\vec K_\varphi\rangle$, such that
\begin{align}\label{eq:solution-dep}\nonumber
&\langle\Psi_{\rm Phys}\left(z,\vec \eta^{(0)}\right)|=\sqrt{2\pi} \langle\Psi_{\rm Phys}\left(z\right)|\vec K_\varphi=\vec \eta^{(0)}\rangle\\
&=\int_0^{\infty} dM\left(\bigotimes_{z(v_j)}\left[\sum_{k_j}\sum_{\mu_j}\psi(M,k_j)\phi(k_j,M;\mu_j)e^{i\gamma \ell_{\rm Pl}^2 \mu_j\eta^{(0)}_j}\langle k_j| \right]\right)\otimes \langle M|.
\end{align} 
Here $\eta_j=k_\varphi(v_j)$ where $\{k_\varphi(v_j)\}$ is the collection of parameters corresponding to the values of the connection $K_\varphi(x)$ restricted to the vertices of $\langle\Psi_{\rm Phys}\left(z\right)|$. Besides we also declare $\{\eta^{(0)}_j\}$ as an ``initial'' collection of parameters codifying the initial choice of Cauchy surface. 

One can easily see that these bras provide a natural definition of the corresponding kets $|\Psi_{\rm Phys}\left(z,\vec \eta^{(0)}\right)\rangle$. In addition, the inner product of two of these states evaluated at the same $z(x)$ and initial collection $\{\eta^{(0)}_j\}$, i.e. $\langle\Psi_{\rm Phys}\left(z,\vec \eta^{(0)}\right)|\Psi_{\rm Phys}\left(z,\vec \eta^{(0)}\right)\rangle$, yields, after some simple calculations, the right hand side of Eq. \eqref{eq:undep-inner}.

Let us notice that there is a map 
\begin{equation}
\hat U\left(\vec \eta,\vec \eta^{(0)}\right): \quad |\Psi_{\rm Phys}\left(z,\vec \eta^{(0)}\right)\rangle\to |\Psi_{\rm Phys}\left(z,\vec \eta\right) \rangle = \hat U\left(\vec \eta,\vec \eta^{(0)}\right)|\Psi_{\rm Phys}\left(z,\vec \eta^{(0)}\right)\rangle,
\end{equation}
that relates states between different slicings $\{\eta^{(0)}_j\}$ and $\{\eta_j\}$. This map would be the analog to the evolution operator in quantum mechanics in the Schr\"odinger picture. It can be decomposed as a product of operators acting on each vertex as 
\begin{equation}
\hat U(\vec \eta,\vec \eta^{(0)})=\prod_{v_j} \hat U_j(\eta_j,\eta^{(0)}_j).
\end{equation}

For the choice of parameter function $K_\varphi(x)$ that we have considered, each of these operators acting on a given vertex in the $\mu_j$-representation is simply
\begin{equation}
\hat U_j(\eta_j,\eta^{(0)}_j)|\mu_j\rangle:= \exp\left\{i\gamma \ell^2_{\rm Pl}\mu_j(\eta_j-\eta^{(0)}_j)\right\}|\mu_j\rangle.
\end{equation}
Then, the mass and the triad operators can be defined as physical observables by means of
\begin{subequations}
\begin{align}\label{eq:solution-dep-Ex}
&\hat M|\Psi_{\rm Phys}\left(z,\vec \eta^{(0)}\right)\rangle=\\\nonumber
&\int_0^{\infty} dM\left(\bigotimes_{z(v_j)}\left[\sum_{k_j}\sum_{\mu_j}M\psi(M,k_j)\phi(k_j,M;\mu_j)e^{i\gamma\ell_{\rm Pl}^2 \mu_j\eta^{(0)}_j}|k_j\rangle \right]\right)\otimes |M\rangle.\\
&\hat E^x(x)|\Psi_{\rm Phys}\left(z,\vec \eta^{(0)}\right)\rangle=\\\nonumber
&\int_0^{\infty} dM\left(\bigotimes_{z(v_j)}\left[\sum_{k_j}\sum_{\mu_j}\gamma\ell_{\rm Pl}^2 k_{{\rm Int}(Nz)}\psi(M,k_j)\phi(k_j,M;\mu_j)e^{i\gamma\ell_{\rm Pl}^2 \mu_j\eta^{(0)}_j}|k_j\rangle \right]\right)\otimes |M\rangle.\\
&\hat E^\varphi(x)|\Psi_{\rm Phys}\left(z,\vec \eta^{(0)}\right)\rangle=\\\nonumber
&\int_0^{\infty} dM\left(\bigotimes_{z(v_j)}\left[\sum_{k_j}\sum_{\mu_j}\gamma\ell_{\rm Pl}^2 \mu_{{\rm Int}(Nz)}\psi(M,k_j)\phi(k_j,M;\mu_j)e^{i\gamma\ell_{\rm Pl}^2 \mu_j\eta^{(0)}_j}|k_j\rangle \right]\right)\otimes |M\rangle.
\end{align} 
\end{subequations}

As we see, this picture presents some advantages with respect to the one given by parametrized observables, since it promotes kinematical phase space functions to physical observables in a simple way. We do not need to previously identify the parametrized observables of the model. This picture only requires the knowledge of the solutions and a suitable inner product. It should be equivalent (however we do not provide a complete proof here) to the picture adopting parametrized observables \cite{gp-lett,gop,gp-semi}. For instance, one can easily see that the observable $\hat O\big(z\big)$ defined in Ref. \cite{gp-lett} corresponds to the action of $\hat E^x(x)$ on these parametrized physical states
\begin{equation}
\hat O(z(x))|\Psi_{\rm Phys}\rangle=\hat E^x(x)|\Psi_{\rm Phys}\left(z\right)\rangle.
\end{equation}
Any  difference that could appear would be due to the way in which classical phase space functions are promoted to physical quantum operators. Then the two strategies would yield analog results {\it up to factor ordering ambiguities}.

\section{Conclusions}\label{sec:concl}

In summary, we have shown that by means of a Dirac quantization of a vacuum spherically symmetric spacetime (with local degrees of freedom) \cite{gp-lett,gop,gp-semi} one can describe its physics either in terms of a Heisenberg-like picture with parametrized Dirac observables as well as with a Schr\"odinger-like picture with parametrized states and the observables constructed out of the kinematical ones restricted to this space of states. More precisely,  since we adopt a loop quantization, we describe the setting in terms of spherically symmetric triads and connections. We then modify the constraint algebra in order to make the scalar constraint Abelian with itself. We review the loop representation for the system. We consider a quantum scalar constraint that is symmetric on ${\cal H}_{\rm kin}$. Its solutions are found by means of group averaging, after assuming that the quantum scalar constraint (at each vertex) is selfadjoint, and the physical inner product is provided. We construct a Schr\"odinger-like picture, where the physical solutions are decomposed in parametrized physical states with a well defined evolution in time (the time function determining the space-like Cauchy surfaces). Besides, the relation between parametrized states under a spatial diffeomorphism is well understood. The basic physical observables of the model (triads) are provided. This physical picture is in agreement with the one provided in previous studies \cite{gp-lett,gop,gp-semi} and has the advantage that kinematical operators are promoted to physical ones in a straightforward manner. 

It is remarkable that most of the aspects studied here have also been considered in simpler models \cite{q-const} (with global degrees of freedom). In all these cases, the group averaging technique plays an essential role in both pictures since it relates kinematical and physical structures (which in general turn out to be inequivalent) through the inner product \eqref{eq:undep-inner}. In addition, the present picture will be very useful in future numerical studies of the dynamics of the present setting, in particular the semiclassical regime, and in the analysis of more general models of spherically symmetric gravity coupled to matter \cite{shell1,shell2} and similar midisuperspace models like Gowdy cosmologies \cite{gowdy-LRS} and Callan--Giddings--Harvey--Strominger (CGHS) dilatonic scenarios \cite{cghs} in loop quantum gravity \cite{gpr1,gpr2}.

\section*{Acknowledgments}

This work has been supported by the grants MICINN/MINECO FIS2011-30145-C03-02 and FIS2014-54800-C2-2-P (Spain), NSF-PHY-1305000 (USA) and Pedeciba (Uruguay).

\end{document}